# Agreement between local and global measurements of the London penetration depth

Thomas M. Lippman[1], Beena Kalisky[1,2], Hyunsoo Kim[3], Makariy A. Tanatar[3], Sergey L. Bud'ko[3], Paul C. Canfield[3], Ruslan Prozorov[3], and Kathryn A. Moler[1]*

*[1]Stanford Institute for Materials and Energy Sciences, SLAC National Accelerator Laboratory, 2575 Sand Hill Road, Menlo Park, CA 94025, USA.*

*[2]Department of Physics, Nano-magnetism Research Center, Institute of Nanotechnology and Advanced Materials, Bar-Ilan University, Ramat-Gan 52900, Israel*

*[3]The Ames Laboratory and Department of Physics and Astronomy, Iowa State University, Ames, Iowa 50011, USA*

*email: kmoler@stanford.edu

Abstract:

Recent measurements of the superconducting penetration depth in $Ba(Fe_{1-x}Co_x)_2As_2$ appeared to disagree on the magnitude and curvature of $\Delta\lambda_{ab}(T)$, even near optimal doping. These measurements were carried out on different samples grown by different groups. To understand the discrepancy, we use scanning SQUID susceptometry and a tunnel diode resonator to measure the penetration depth in a single sample. The penetration depth observed by the two techniques is identical with no adjustments. We conclude that any discrepancies arise from differences between samples, either in growth or crystal preparation.

Measurements of the London penetration depth are an important part of determining the symmetry of the order parameter in unconventional superconductors.  The temperature-induced change in the penetration depth, $\Delta\lambda_{ab}(T)$, is a sensitive measure of low-energy quasiparticles in the superconducting state.  This makes it a powerful probe of the magnitude of the energy gap $\Delta(k)$, although the multi-band nature of the pnictides complicates the picture.[1,2]

Further complicating the picture is the quantitative disagreement between penetration depth measurements on similar samples.  Specifically, early tunnel diode resonator (TDR) measurements of the $Ba(Fe_{1-x}Co_x)_2As_2$ system close to optimal doping (x = 0.074) reported $\Delta\lambda_{ab}(T) \sim T^n$ with n = 2.4 and $\Delta\lambda_{ab}(8\ K)$ of around 100 nm.[3]  Subsequent TDR measurements on a similar sample from a different batch yielded n = 2.8 and a dramatically smaller $\Delta\lambda_{ab}(8\ K)$ of 13 nm.[4]  A sample from the same batch was also measured using low-energy muon spin rotation and microwave cavity perturbation, and the results are consistent with the TDR data.[5]  Measurements on a third sample with local magnetic probes had n = 3.0 and $\Delta\lambda_{ab}(8\ K) = 5.5$ nm.[6]  This third sample had a slightly different composition (x = 0.07) and was grown by a different group.

This discrepancy could be due to differences between measurement techniques or variations in electronic properties among samples.  By performing different measurements *on the same sample*, we are able to rule out the former possibility.

Single crystals of $Ba(Fe_{1-x}Co_x)_2As_2$ were grown by a self-flux method as described elsewhere.[7]  The superconducting $T_c$ is 22.5 K, and wavelength dispersive spectroscopy shows that x = 0.074, making this an optimally doped sample.  In Ref. 4 Kim and collaborators used a TDR to obtain global measurements of the temperature-induced change of the in-plane London penetration depth.  The technique is described in Ref. 4 and references therein.  In this work, we report local penetration depth measurements performed on the same sample.  The sample was primarily stored in a desiccator, but was exposed to atmosphere during shipment and cooldown.  No additional cleaving was performed and the sample was measured as received after TDR measurements, which occurred 30 months prior. We use a scanning superconducting quantum interference device susceptometer (SSS) to measure the local diamagnetic response of the sample at three locations, as described previously.[8]  The temperature-induced changes in diamagnetic response can be converted to changes in penetration depth by the scanner calibration constant, which dominates our 7% systematic uncertainty in $\Delta\lambda_{ab}(T)$.[8]  By repeating the SSS measurement at different locations on the same sample, we can observe any large scale inhomogeneity (>100 $\mu$m) in the sample, if it is present.  To account for a slow, irreproducible drift in the measurement, we report data that re-traces itself on sweeping the temperature up and down.

We compare the local and global penetration depth measurements in Fig 1.  The local SSS measurements were done at three locations on the sample surface, shown as colored marks in Fig 1a.  The three locations agree well with one another, and with the global TDR measurement, shown in Fig 1b.  This agreement between global and local measurements of the penetration depth is the main message of this Brief Report.  We note that each group has used its own calibration procedure to convert the raw data into the penetration depth, and that there are no adjustments to or rescaling of the data in Fib. 1b.

We also plot in Fig. 1b local penetration depth data from a different sample, previously reported in Ref. 6. This sample is grown with a slightly different procedure[9] and has a composition (x = 0.07) and transition temperature ($T_c$ = 22.4 K) that are essentially identical to the present sample (x = 0.074 and $T_c$ = 22.5 K). The data from Ref. 6 has a smaller overall magnitude $\Delta\lambda_{ab}$(20 K), by a factor of roughly 2.5, and is flatter at low temperature than the data on the current sample.

By measuring the same sample with a global (TDR) and a local (SSS) technique we have shown that there is no disagreement between the two.  It then stands to reason that previously reported discrepancies in $\Delta\lambda_{ab}$(T) are due to variation between samples, due either to differences in growth or differences in crystal preparation (e.g. cutting, cleaving, deforming, etc.), not differences among experimental methods.

The different behavior of $\Delta\lambda_{ab}$(T) in the two samples is striking, given their apparent similarity.  Both are optimally doped, with no structural or magnetic phase transition present.  The current sample, from the Ames group, has a doping of x = 0.074 and a resistive $T_c$ of 22.5 K.  The doping level is determined by wavelength dispersive spectroscopy, and has an uncertainty of 0.003.[7]  By comparison, the sample from the Stanford group used in Ref. 6 has a doping of x = 0.07 and a resistive $T_c$ of 22.4 K.  The doping level is inferred from electron microprobe analysis on similar samples, and has an uncertainty of 0.0015.[9]

There is qualitative agreement of the penetration depth in the two samples. Both datasets are inconsistent with a single weak-coupling BCS gap, but can be explained by one or more fully gapped two-band scenarios.  But a quantitative comparison of the two suggests that there are approximately twice as many quasiparticles at low temperature in the Ames sample as in the Stanford sample. This difference could arise from different amounts of pair-breaking interband scattering, raising the question of why the scattering rates should be so different. Any explanation based on surface quality seems unlikely, because the supercurrent response occurs over a few hundred nanometers, so we expect it to be less sensitive to surface degradation than probes like scanning tunneling microscopy or photoemission that are sensitive only to the top few atomic layers.  In addition, all of the samples considered here are exposed to atmosphere after cleaving, so the surface quality should not be radically different. Careful physical and chemical investigation of the nominally identical samples produced by various groups might give an answer to the differences. However, the goal of this Brief Report is to show quantitative agreement between very different techniques to measure London penetration depth.

We have shown that there is agreement between a global (TDR) and a local (SSS) technique for measuring the temperature dependence of the penetration depth on the same sample when the response across the sample is homogeneous and the sample surface has few terraces.  This work complements the recently demonstrated agreement between TDR, μSR, and microwave methods of measuring the temperature dependence of the penetration depth.[5]

Acknowledgments:
Work at Stanford was supported by the Department of Energy, Office of Basic Energy Sciences, Division of Materials Sciences and Engineering, under contract DE-AC02-76SF00515.  Work at the Ames Laboratory was supported by the Department of Energy, Basic Energy Sciences, Division of Materials Sciences and Engineering under Contract No. DE-AC02-07CH11358.

Bibliography:
1 A. B. Vorontsov, M. G. Vavilov and A. V. Chubukov, Physical Review B (Condensed Matter and Materials Physics) 79, 140507-4 (2009).
2 R. Prozorov and V G Kogan, Reports on Progress in Physics 74, 124505 (2011).
3 R. T. Gordon, N. Ni, C. Martin, M. A. Tanatar, M. D. Vannette, H. Kim, G. D. Samolyuk, J. Schmalian, S. Nandi, A. Kreyssig, A. I. Goldman, J. Q. Yan, S. L. Bud'ko, P. C. Canfield and R. Prozorov, Physical Review Letters 102, 127004-4 (2009).
4 H. Kim, R. T. Gordon, M. A. Tanatar, J. Hua, U. Welp, W. K. Kwok, N. Ni, S. L. Bud'ko, P. C. Canfield, A. B. Vorontsov and R. Prozorov, Physical Review B 82, 060518 (2010).
5 Oren Ofer, J C Baglo, M D Hossain, R F Kiefl, W N Hardy, A Thaler, H Kim, M A Tanatar, P C Canfield, R Prozorov, G M Luke, E Morenzoni, H Saadaoui, A Suter, T Prokscha, B M Wojek and Z Salman, Phys. Rev. B 85, 060506 (2012).
6 L. Luan, T. M. Lippman, C. W. Hicks, J. A. Bert, O. M. Auslaender, Jiun-Haw Chu, J. G. Analytis, I. R. Fisher and K. A. Moler, Physical Review Letters 106, 067001 (2011).
7 N Ni, M E Tillman, J -Q Yan, A Kracher, S T Hannahs, S L Bud'ko and P C Canfield, Phys. Rev. B 78, 214515 (2008).
8 C. W. Hicks, T. M. Lippman, M. E. Huber, J. G. Analytis, Jiun-Haw Chu, A. S. Erickson, I. R. Fisher and K. A. Moler, Physical Review Letters 103, 127003 (2009).
9 Jiun-Haw Chu, J. G. Analytis, C. Kucharczyk and I. R. Fisher, Physical Review B 79, 014506 (2009).

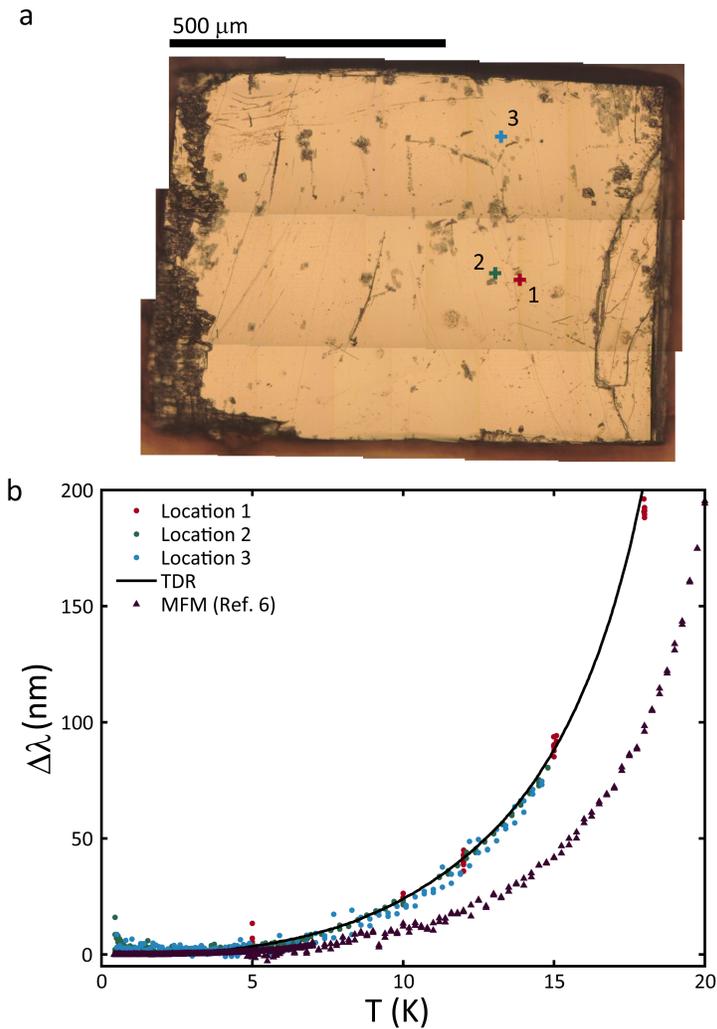

Figure 1: Comparison of penetration depth data taken by scanning SQUID (colored marks) and tunnel diode resonator (black line, Ref. 4) on the same sample, 7.4% doping. We find good agreement among the three locations where we measured the penetration depth with scanning SQUID, and also between the SQUID and tunnel diode resonator measurements. For comparison we also plot previous MFM and scanning SQUID results on a similar sample. The discrepancy between the data shows that the pnictides are exquisitely sensitive to subtle details of sample growth and preparation.